\documentclass[12pt]{article}
\usepackage{graphicx}

\def \hf{\frac{1}{2}}

\def \bea{\begin{eqnarray}}
\def \beq{\begin{equation}}

\def \eea{\end{eqnarray}}
\def \eeq{\end{equation}}

\def \({\left(}
\def \){\right)}
\def \[{\left[}
\def \]{\right]}

\def \s{\sqrt{2}}

\begin{document}
\rightline{TECHNION-PH-14-11}
\rightline{July 2014}
\rightline{arXiv:1407.7374}
\vskip 10mm
\centerline{\bf NEW PHYSICS IN SINGLY CABIBBO-SUPPRESSED $D$ DECAYS}
\bigskip
\centerline{Michael Gronau}
\medskip
\centerline{\it Physics Department, Technion -- Israel Institute of Technology}
\centerline{\it Haifa 3200, Israel}
\bigskip
\begin{quote}
A U-spin relation among four ratios of amplitudes for $D^0 \to \pi^+K^-$,
$K^+\pi^-$, $K^+K^-, \pi^+\pi^-$, including first, second and third order U-spin breaking,  
has been derived recently with a precision of $10^{-3}$. We study effects of 
new $|\Delta C|=1$ operators on this relation. We find that it is not affected by 
U-spin scalar operators, including QCD penguin and chromomagnetic dipole 
operators occurring in supersymmetric and extra-dimensional models. The relation
is modified by new $U=1$ operators with a sensitivity 
of a few percent 
characteristic of second order U-spin breaking. Combining this relation with CP 
asymmetries in $D^0\to K^+K^-, \pi^+\pi^-$ leads to a more solid constraint on $U=1$ 
operators than from asymmetries alone.
 \end{quote}
\bigskip

\section{Introduction\label{sec:introduction}}

Indirect evidence for new physics in processes involving charmed mesons may be 
obtained in two ways: 
\begin{itemize}
\item Very rare processes could be measured with rates that exceed unambiguously 
predictions within the Standard Model. 
\item Very precise amplitude relations predicted within the Standard Model could be 
violated experimentally.
\end{itemize}
Two properties of  charmed mesons, $D^0$--$\bar D^0$ mixing and CP violation in singly Cabibbo-suppressed (SCS) $D$ meson decays, have been suggested as potential cases
for the first scenario~\cite{Bianco:2003vb,Golowich:2007ka,Grossman:2006jg}. 
Recently we derived a very precise nonlinear relation among four ratios of amplitudes for 
$D^0 \to \pi^+K^-, K^+\pi^-, K^+K^-, \pi^+\pi^-$~\cite{Gronau:2013xba}, valid up to fourth order U-spin breaking. 
While precise amplitude relations have already been proposed for hadronic $B$ meson 
decays (see for instance Ref.\,\cite{Gronau:1990ka}), this particular relation provides a first 
case for the second scenario in hadronic $D$ decays.

Measurements of the difference between CP asymmetries  in $D^0\to K^+K^-$ and
$D^0\to\pi^+\pi^-$ of order $10^{-3}$~\cite{Aaltonen:2011se,Aaij:2011in,HFAG} have been 
shown to be consistent with Standard Model estimates~\cite{Brod:2011re}. This has been 
used to obtain model-dependent constraints on new $|\Delta C|=1$ operators occurring in a number of models~\cite{Isidori:2011qw}. The amplitude relation derived in 
Ref.~\cite{Gronau:2013xba}, involving a precision of $10^{-3}$, has been 
shown to agree with experiment at this same high accuracy. The purpose of this 
letter is to study the possibility of using this excellent agreement for obtaining
model-independent constraints on new $|\Delta C|=1$ operators.

The proof in~\cite{Gronau:2013xba} of the nonlinear amplitude relation is based largely on the fact that the charm-changing weak hamiltonian transforms as a U-spin triplet~\cite{Feldmann:2012js,Gronau:2012kq}. 
Thus we will distinguish between two classes of models involving new $|\Delta C|=1$ operators behaving distinctly under U-spin. In the first case we will assume these operators to transform like 
U-spin scalars. This rather broad class of models includes supersymmetric and extra-dimensional models involving new QCD penguin and chromomagnetic dipole operators. Constraints on such models from CP asymmetries in SCS D decays and from $D^0$-$\bar D^0$ mixing have been studied in Refs.~\cite{Grossman:2006jg} and~\cite{DaRold:2012sz}. A second class of models includes new $|\Delta C|=1$ operators transforming like $U=1, U_3=0$. Constraints on such operators from CP asymmetries in SCS decays have been discussed in Ref.~\cite{Hiller}.
Other probes for new physics have been suggested in Ref.\,\cite{Grossman:2012eb}
in terms of isospin sum rules for CP asymmetries in SCS $D$ decays. 

Section~\ref{sec:SM} summarizes briefly arguments used in 
Ref.~\cite{Gronau:2013xba} leading to a precise nonlinear relation among four ratios 
of amplitudes for $D^0 \to \pi^+K^-, K^+\pi^-, K^+K^-, \pi^+\pi^-$. In Sections~\ref{sec:U=0} 
and~\ref{sec:U=1} we study separately contributions of new $U=0$ and $U=1, U_3=0$ 
operators potentially modifying this relation. Section~\ref{sec:CP asymmetries} contains a
discussion of CP asymmetries in $D^0\to K^+K^-$ and $D^0 \to \pi^+\pi^-$ in these two 
classes of models. Conclusions are given in Section~\ref{sec:conclusion}.  

\section{Precise amplitude relation in the Standard Model\label{sec:SM}}

Hadronic weak decays of charmed mesons are conveniently studied using U-spin 
symmetry under which the quark pair $(d,s)$ transforms like a doublet.
The effective Hamiltonian operators for Cabibbo-favored (CF), singly Cabibbo-suppressed (SCS) and doubly Cabibbo-suppressed charm decays transform like three components $U_3=-1, 0, +1$
of $U=1$, excluding corresponding CKM factors, $\cos^2\theta_C, -\cos\theta_C\sin\theta_C, \sin^2\theta_C$. 

We neglect a term in the SCS Hamiltonian proportional to a tiny CKM factor $V^*_{cb}V_{ub}$, where $|V^*_{cb}V_{ub}|/\cos\theta_C\sin\theta_C \simeq 0.7\times 10^{-3}$~\cite{PDG}.
This term would be responsible for CP asymmetries of this order in $D^0\to K^+K^-$ and 
$D^0\to\pi^+\pi^-$. Since the measured CP asymmetries are at most of this order~\cite{HFAG}, while our approximation for amplitudes involves uncertainties of this same order, we will neglect
CP asymmetries also in the next two sections discussing new physics, returning to discuss them
in Section \ref{sec:CP asymmetries}.

The $D^0$ is a U-spin singlet,  the states $-|\pi^+K^-\rangle, \frac{1}{\s}|K^+K^--\pi^+\pi^-\rangle, |K^+\pi^-\rangle$ are three components $|U_3=-1, 0, +1\rangle$ of $|U=1\rangle$ while
$\frac{1}{\s}|K^+K^- + \pi^+\pi^-\rangle$ is a singlet. 
The matrix element of the $U=1$ Hamiltonian vanishes for the latter state, 
$\langle K^+K^-|{\cal H}_{\rm eff}|D^0\rangle = - \langle \pi^+\pi^-|{\cal H}_{\rm eff}|D^0\rangle$.
Thus in the U-spin symmetry limit the four amplitudes for $D^0 \to \pi^+K^-, K^+K^-, \pi^+\pi^-, K^+\pi^-$ are given by a common $U=1$ amplitude  $A$. Consequently the multiple ratio of the four decay amplitudes is given 
by ratios of CKM factors~\cite{Kingsley:1975fe}: 
\bea
& & A(D^0\to \pi^+K^-) : A(D^0\to K^+K^-) : A(D^0\to \pi^+\pi^-) : A(D^0 \to K^+\pi^-) 
\nonumber\\
&  & 
= \cos^2\theta_C : \cos\theta_C\sin\theta_C : -\cos\theta_C\sin\theta_C : -\sin^2\theta_C~.
\eea

U-spin breaking in amplitudes is treated perturbatively in terms of two distinct parameters
proportional to $(m_s - m_d)/\Lambda_{\rm QCD}$, taken separately for $D^0\to \pi^+K^-, 
K^+\pi^-$ and $D^0 \to K^+K^-, \pi^+\pi^-$. Symmetry breaking of order $k$ in
an amplitude $\langle f|{\cal H}_W|D^0\rangle$ is obtained by introducing in ${\cal H}_W$
or in $|f\rangle$ $k$ powers of a quark mass-difference operator $\bar s s- \bar d d$ behaving 
like $U=1, U_3=0$. The decays $D^0 \to K^+K^-, \pi^+\pi^-$ obtain also a first order U-spin breaking correction from a $U=0$ penguin operator. This and a simple sign property for U-spin
addition, $(1, -1;n, 0|1, -1) = (-1)^n(1, 1;n, 0|1, 1)$, lead to the following two properties of U-spin breaking~\cite{Gronau:2013xba}:
\begin{itemize}
\item Symmetry breaking effects in $D^0\to \pi^+K^-, K^+\pi^-$ and $D^0 \to K^+K^-, \pi^+\pi^-$
are described by two different parameters, to be denoted $\epsilon_1$ and $\epsilon_2$, 
respectively.
\item In each one of these two pairs of processes U-spin breaking corrections of even (odd) 
order have equal magnitudes and same (opposite) signs. 
\end{itemize} 
Expanding up to third order U-spin breaking one has
\bea\label{Amp}
A(D^0 \to \pi^+K^-) & = & \cos^2\theta_CA[1 - \epsilon_1 + a_1(\epsilon_1)^2
- a'_1(\epsilon_1)^3]~, 
\nonumber\\
A(D^0\to K^+\pi^-) & = & -\sin^2\theta_CA[1 + \epsilon_1 + a_1(\epsilon_1)^2
+ a'_1(\epsilon_1)^3]~, 
\nonumber\\
A(D^0 \to K^+K^-) & = & \cos\theta_C\sin\theta_CA[1 + \epsilon_2 + a_2(\epsilon_2)^2
+ a'_2(\epsilon_2)^3]~, \nonumber\\
A(D^0 \to \pi^+\pi^-) & = & -\cos\theta_C\sin\theta_CA[1 - \epsilon_2 + a_2(\epsilon_2)^2
- a'_2(\epsilon_2)^3]~,
\eea
where $a_{1,2} \sim a'_{1,2} \sim 1$.

Defining four independent ratios of amplitudes $R_i, (i=1,2,3,4)$ one finds:
\bea\label{Ri}
R_1  & \equiv &  \frac{|A(D^0 \to K^+\pi^-)|}{|A(D^0 \to \pi^+K^-)|\tan^2\theta_C}
= 1 + 2[{\rm Re}\,\epsilon_1 + ({\rm Re}\,\epsilon_1)^2] = 1.118 \pm 0.014~,
\nonumber\\
R_2 & \equiv & \frac{|A(D^0 \to K^+K^-)|}{|A(D^0\to \pi^+\pi^-)|} =
1 + 2[{\rm Re}\,\epsilon_2 + ({\rm Re}\,\epsilon_2)^2] = 1.814 \pm 0.018~,
\nonumber\\
R_3 & \equiv & \frac{|A(D^0 \to K^+K^-)| + |A(D^0\to \pi^+\pi^-)|}
{|A(D^0 \to \pi^+K^-)|\tan\theta_C + |A(D^0\to K^+\pi^-)|\tan^{-1}\theta_C} 
\nonumber\\
& = & 1 + \hf[({\rm Im}\,\epsilon_2)^2 - ({\rm Im}\,\epsilon_1)^2] + 
{\rm Re}\,[a_2(\epsilon_2)^2 - a_1(\epsilon_1)^2] = 1.056 \pm 0.008~,
\nonumber\\
R_4 & \equiv & \sqrt{\frac{|A(D^0 \to K^+K^-)||A(D^0\to \pi^+\pi^-)|}
{|A(D^0\to \pi^+K^-)||A(D^0\to K^+\pi^-)|}} 
= 1 + \hf[({\rm Im}\,\epsilon_2)^2 - ({\rm Im}\,\epsilon_1)^2]
\nonumber\\
& + & {\rm Re}\,[a_2(\epsilon_2)^2 - a_1(\epsilon_1)^2]  
- \hf[({\rm Re}\,\epsilon_2)^2 - ({\rm Re}\,\epsilon_1)^2]
= 1.012 \pm 0.007~.
\eea
$R_1$ and $R_2$ involve additional third order terms while corrections to $R_3$ 
and $R_4$ start at fourth order. [The last term of Eq,\,(21) in \cite{Gronau:2013xba} 
should be of fourth order.]
In the above we used the following third order expansion:
\bea
|1 \pm \epsilon + a\epsilon^2 \pm a'\epsilon^3| = 1 & \pm & {\rm Re}\,\epsilon + 
\hf({\rm Im}\,\epsilon)^2 + {\rm Re}(a\epsilon^2) \pm {\rm Re}(a'\epsilon^3)
\nonumber\\
& \mp & \hf{\rm Re}\,\epsilon\,({\rm Im}\,\epsilon)^2 
\mp {\rm Im}\,\epsilon\,{\rm Im}(a\epsilon^2)~.
\eea 
\begin{table}[h]
\caption{Amplitudes in units of 
$10^{-1}({\rm GeV}/c)^{-1/2}$ for $D^0$ decays to pairs involving a charged pion 
and kaon.\label{tab:A}} 
\begin{center}
\begin{tabular}{c c } \hline \hline
Decay mode  & $|A|=\sqrt{{\cal B}/p^*}$ \\ \hline
 $D^0\to \pi^+K^-$  &  $2.1228$  \\ 
 $D^0 \to K^+\pi^- $ &  $0.1268 \pm  0.0012$ \\
 $D^0 \to K^+K^-$ & $0.7076 \pm 0.0052$ \\ 
 $D^0 \to \pi^+\pi^-$ & $0.3900 \pm 0.0027$ \\ 
 \hline \hline
\end{tabular}
\end{center}
\end{table}

Numerical values on the right hand sides of (\ref{Ri})  have been obtained using  
$\tan\theta_C=0.2312\pm 0.0009$ and branching 
fractions for $D^0\to K^+\pi^-, K^+K^-, \pi^+\pi^-$ measured relative to 
$D^0 \to \pi^+K^-$~\cite{PDG}. Table \ref{tab:A} quotes magnitudes of the four amplitudes 
defined by $|A| \equiv \sqrt{{\cal B}/p^*}$. Note that the first amplitude involves no error as 
the three others are measured relative to its magnitude.

The second order expressions in (\ref{Ri}) imply ${\rm Re}\epsilon_1 =0.056 \pm 0.006, {\rm Re}\epsilon_2 = 0.311 \pm 0.006$, and a nonlinear relation among these four ratios which holds up
to fourth order U-spin breaking:
\beq\label{NLrel}
\Delta R \equiv R_3 - R_4 + \frac{1}{8}\left[(\sqrt{2R_1-1} - 1)^2 - (\sqrt{2R_2 - 1} -1)^2\right]=0~.
\eeq
This relation is satisfied extremely well by current experiments for which one finds
\beq
\Delta R_{\rm exp} = -0.003 \pm 0.002~.
\eeq
We have neglected in Eq.\,(\ref{NLrel}) a fourth order U-spin breaking correction and an 
isospin breaking term suppressed also by U-spin breaking. This leads to an uncertainty 
of order $10^{-3}$~\cite{Gronau:2013xba}.

\section{Contributions of a new $U=0$ operator\label{sec:U=0}}

A new $U=0$ operator does not contribute to $D^0 \to \pi^\pm K^\mp$ where final states
have $U_3=\mp 1$. This is true in the U-spin symmetry limit and also when including U-spin 
breaking of arbitrary order.  We will assume that contributions of this operator in 
$D^0\to K^+K^-$ and $D^0\to \pi^+\pi^-$ are subleading, namely of order $\epsilon_2$ 
or smaller, as are the U-spin breaking
contributions of a $U=0$ penguin operator occurring in the CKM
 framework~\cite{Gronau:2013xba}. In these two processes the new $U=0$ operator has 
equal contributions in the U-spin symmetry limit and first order U-spin breaking terms of 
equal magnitudes and opposite signs. Normalizing these two contributions by the $U=1$ amplitude we denote them by $\cos\theta_C\sin\theta_CAn$ and 
$\cos\theta_C\sin\theta_CAn\epsilon$, respectively:
\bea\label{amp0}
A(D^0 \to K^+K^-) & = & \cos\theta_C\sin\theta_CA[1 + \epsilon_2 + a_2(\epsilon_2)^2 
+ a'_2(\epsilon_2)^3 + n + n\epsilon]~,
\nonumber\\
A(D^0 \to \pi^+\pi^-) & = & -\cos\theta_C\sin\theta_CA[1 - \epsilon_2 + a_2(\epsilon_2)^2 
- a'_2(\epsilon_2)^3 - n + n\epsilon]~.
\eea  
Here we wish to study the effects of these new terms on (\ref{Ri}) and (\ref{NLrel}).

Expanding ratios of amplitudes up to and including terms of second order in $n$ and 
in the U-spin breaking parameters $\epsilon_1, \epsilon_2, \epsilon$, one obtains
\bea\label{RiU=0}
R_1 & = & 1 + 2[{\rm Re}\,\epsilon_1 + ({\rm Re}\,\epsilon_1)^2]~,
\nonumber\\
R_2 & = & 1 + 2\left[{\rm Re}\,(\epsilon_2 + n) + [{\rm Re}\,(\epsilon_2+n)]^2\right]~,
\nonumber\\
R_3 & = & 1 + \hf[({\rm Im}\,(\epsilon_2+n))^2 - ({\rm Im}\,\epsilon_1)^2] + 
{\rm Re}\,[a_2(\epsilon_2)^2 - a_1(\epsilon_1)^2 + n\epsilon] 
\nonumber\\
R_4 & = & 1 
+  \hf[({\rm Im}\,(\epsilon_2+n))^2 - ({\rm Im}\,\epsilon_1)^2] + 
{\rm Re}\,[a_2(\epsilon_2)^2 - a_1(\epsilon_1)^2 +n\epsilon] 
\nonumber\\
& &
 ~~- \hf[({\rm Re}\,(\epsilon_2+n))^2 - ({\rm Re}\,\epsilon_1)^2]~.
\eea
These results correspond to a substituting $\epsilon_2 \to \epsilon_2 + n$ 
and $a_2(\epsilon_2)^2 \to a_2(\epsilon_2)^2 + n\epsilon$ in (\ref{Ri}). 

Thus the ratio $R_2$ involves a term which is first order in the $U=0$ amplitude. It implies
${\rm Re}(\epsilon_2 + n) = 0.311 \pm 0.006$.
The ratios $R_3$ and $R_4$ include identical second order terms depending on the 
$U=0$ amplitude that cancel in their difference occurring in $\Delta R$. Since 
Eqs.\,(\ref{RiU=0}) have the same structure as Eqs.\,(\ref{Ri}), with a mere substitution,
$\epsilon_2 \to \epsilon_2 + n, a_2(\epsilon_2)^2 \to a_2(\epsilon_2)^2 + n\epsilon$,
the nonlinear relation (\ref{NLrel}) still holds. That is, {\em Eq.\,(\ref{NLrel}) is unaffected 
by arbitrary new $U=0$ operators and cannot be used to constrain such operators}.

\section{Contributions of a new $U=1, U_3=0$ operator\label{sec:U=1}}

Consider now models with new $U=1, U_3=0$ operators. The effect of such 
operators on (\ref{Amp}) is to modify the overall factor in the amplitudes for 
$D^0\to K^+K^-, \pi^+\pi^-,$ and to replace the U-spin breaking parameter $\epsilon_2$
by a new parameter $\epsilon'_2$, corresponding to U-spin breaking in the total $U=1, 
U_3=0$ amplitude. Using a parameter $n$ to normalize the new amplitude by the 
U-spin invariant amplitude $A$, one has
\bea\label{KKpipi} 
A(D^0 \to K^+K^-) & = & \cos\theta_C\sin\theta_CA(1 + n)[1+ \epsilon'_2 + 
a_2(\epsilon'_2)^2 + a'_2(\epsilon'_2)^3]~,
\nonumber\\
A(D^0 \to \pi^+\pi^-) & = & -\cos\theta_C\sin\theta_CA(1 + n)[1 - \epsilon'_2 + 
a_2(\epsilon'_2)^2 - a'_2(\epsilon'_2)^3]~.
\eea

Expanding the four ratios of amplitudes up to second order, we note that
$R_1$ and $R_2$ are essentially unaffected relative to (\ref{Ri}) while  
$R_3$ and $R_4$ obtain an overall factor $|1+n|$:
\bea
R_1  & = & 1 + 2[{\rm Re}\,\epsilon_1 + ({\rm Re}\,\epsilon_1)^2]~,
\nonumber\\
R_2 & = & 1 + 2[{\rm Re}\,\epsilon'_2 + ({\rm Re}\,\epsilon'_2)^2]~~~~
{\rm implying}~{\rm Re}\,\epsilon'_2=0.311\pm 0.006~,
\nonumber\\
R_3 & = & |1 + n|\left[1 + \hf[({\rm Im}\,\epsilon'_2)^2 - ({\rm Im}\,\epsilon_1)^2] + 
{\rm Re}\,[a'_2(\epsilon'_2)^2 - a_1(\epsilon_1)^2]\right]~,
\nonumber\\
R_4 & = & |1+n|\left[ 1 + \hf[({\rm Im}\,\epsilon'_2)^2 - ({\rm Im}\,\epsilon_1)^2]
+  {\rm Re}\,[a'_2(\epsilon'_2)^2 - a_1(\epsilon_1)^2] \right. 
\nonumber\\
~~~~~~& & \left.- \hf[({\rm Re}\,\epsilon'_2)^2 - ({\rm Re}\,\epsilon_1)^2]\right]~.
\eea

Thus the relation (\ref{NLrel}) is now modified to
\beq
R_3 - R_4 + \frac{1}{8}|1+n|\left[(\sqrt{2R_1-1} - 1)^2 - (\sqrt{2R_2 - 1} -1)^2\right] = 0~,
\eeq
leading to the following constraint on the complex parameter $n$ representing a new $U=1,
U_3=0$ amplitude: 
\beq\label{1+n}
|1+n| = \frac{R_3 - R_4}{\frac{1}{8}[(\sqrt{2R_2-1} - 1)^2 - (\sqrt{2R_1 - 1} -1)^2]}~.
\eeq
Using values of amplitudes given in Table \ref{tab:A} we calculate
\beq\label{num1+n}
|1 + n| \simeq 1 + {\rm Re}\, n= (0.95 \pm 0.02)
\left[1 + {\cal O}\left((\epsilon_1)^2, (\epsilon'_2)^2\right)\right]~.
\eeq
Fourth order U-spin breaking corrections have been neglected in the numerator and 
denominator of (\ref{1+n}), which by themselves are both of second order. Therefore  
Eq.\,(\ref{num1+n}) is valid up to second order U-spin breaking. Second order terms 
in $R_3$ and $R_4$ in (\ref{Ri}) have been shown to be between one and five percent.
Adding in quadrature this uncertainty and the experimental error in (\ref{num1+n}) we obtain
\beq\label{Ren}
{\rm Re}\,n = -0.05\pm 0.05~.
\eeq.

\section{CP asymmetries in $D^0\to K^+K^-, \pi^+\pi^-$\label{sec:CP asymmetries}}

In the preceding sections we have neglected CP asymmetries in $D^0\to K^+K^-, \pi^+\pi^-$,
which are expected to be at most of order $10^{-3}$ in the CKM framework in agreement with
experiments~\cite{HFAG}. We have also neglected these asymmetries in the presence of new operators, consistent with neglecting
fourth order U-spin breaking which introduces uncertainties of this order.
In general, hadronic matrix elements of these new operators may involve a CP-violating 
phase $\phi$ and a strong phase $\delta$, $n = |n|e^{i\delta}e^{i\phi}$.
Consequently the two processes, $D^0\to K^+K^-, \pi^+\pi^-$ acquire nonzero CP asymmetries, 
\beq\label{ACP}
A_{\rm CP}(D^0 \to f) \equiv 
\frac{|A(D^0 \to f)|^2 - |A(\bar D^0 \to \bar f)}{|A(D^0 \to f)|^2 - |A(\bar D^0 \to \bar f)}~,
\eeq
which are proportional to $|n|\sin\delta\sin\phi$. The ratios of amplitudes $R_i$ in Eqs.(\ref{Ri}) 
are now defined in terms of CP-averaged amplitudes
\beq
|A(D^0 \to f)|_{\rm CPav} \equiv \sqrt{\hf\left[|A(D^0 \to f)|^2 + |A(\bar D^0\to \bar f)|^2\right]}~.
\eeq
These amplitudes involve a term $|n|\cos\delta\cos\phi$ instead of
${\rm Re}\,n$ occuring in the above discussion neglecting CP asymmetries. We now summarize the situation of $\Delta R$ and $A_{\rm CP}$ defined in (\ref{NLrel}) and (\ref{ACP}) in the presence of a new CP violating phase $\phi$. 
 \begin{itemize}
 \item For new $U=0$ operators the ratio $R_2$ and the difference $R_3- R_4$ obtain 
 expressions as in (\ref{Ri}) where one substitutes ${\rm Re}\,\epsilon_2 \to {\rm Re}\,\epsilon_2 
 + |n|\cos\delta\cos\phi$. Thus Eq.~(\ref{NLrel}) holds also for $\phi\ne 0$.
The two CP asymmetries are equal in the U-spin symmetry approximation and have opposite 
signs
(as are small CKM asymmetries \cite{Feldmann:2012js}),    
\beq\label{ACPU0}
A_{\rm CP}(D^0\to \pi^+\pi^-)\simeq -A_{\rm CP}(D^0 \to K^+K^-) \simeq 2|n|\sin\delta\sin\phi~.
\eeq
\item For new $U=1$ operators Eq.\,(\ref{Ren}) becomes 
\beq\label{1+nU1}
|n|\cos\delta\cos\phi = -0.05 \pm 0.05~.
\eeq
The two asymmetries are exactly equal and have the same sign,
\beq\label{ACPU1} 
A_{\rm CP}(D^0\to \pi^+\pi^-) = A_{\rm CP}(D^0 \to K^+K^-) \simeq -2|n|\sin\delta\sin\phi~.
\eeq
\end{itemize}
Experimental constraints of CP asymmetries on $|n|$ implied by (\ref{ACPU0}) or (\ref{ACPU1}) depend on unknown values of $\delta$ and $\phi$. An uncertainty in the factor $\sin\delta\sin\phi$ is intrinsic in all earlier work studying constraints on new physics from CP asymmetries in singly 
Cabibbo-suppressed $D$ decays~\cite{Grossman:2006jg,Isidori:2011qw}.
One often assumes $\sin\delta\sin\phi\sim 1$, thereby obtaining the strongest possible 
constraint on $|n|$ of order $10^{-3}$. However, the constraint becomes much weaker for 
small values of $\phi$ or $\delta$ and no constraint is obtained for $\phi=0$ or $\delta =0$. 

Our other restriction (\ref{1+nU1}) involves the factor $\cos\delta\cos\phi$ which is complementary 
to $\sin\delta\sin\phi$, becoming maximal for $\phi=0, \delta=0$. Thus combining (\ref{1+nU1})  and 
(\ref{ACPU1}) 
leads to a more robust constraint on $|n|$ than obtained by using merely the two CP 
asymmetries.

We note in passing that the new amplitude $n$ is absorbed into the definitions of 
$A(D^0\to K^+K^-)$ and $A(D^0 \to \pi^+\pi^-)$ in (\ref{KKpipi}).
Therefore it does not affect the contributions of these amplitudes to the $D^0-\bar D^0$ 
mixing parameter $y\equiv \Delta\Gamma/2\Gamma$. These contributions are only a small 
fraction of the measured mixing parameter~\cite{Gronau:2012kq}.
  
\section{Conclusion\label{sec:conclusion}} 
  
We have studied the effects of new physics operators on a precise U-spin relation 
for $D^0$ decays to pairs involving a charged pion or kaon. We have shown that 
this relation is unaffected by new $U=0$ operators, while its sensitivity to new $U=1$ 
operators is at a level 
of a few percent characteristic of second order U-spin breaking.  
The two classes of models involving $U=0$ and $U=1$ operators may be distinguished 
by the relative sign of CP asymmetries in $D^0 \to K^+K^-$ and $D^0 \to \pi^+\pi^-$.

\end{document}